\documentclass{nature}
\usepackage{color}
\usepackage{amssymb}
\usepackage{graphicx}
\usepackage[normalem]{ulem}
\usepackage{multibib}
\newcites{main,methods}{References,Additional References}
\usepackage{epstopdf}
\makeatletter
\let\saved@includegraphics\includegraphics
\AtBeginDocument{\let\includegraphics\saved@includegraphics}
\renewenvironment*{figure}{
	\AtEndDocument{
    \newpage
    \@float{figure}}
    }{\end@float}
\renewenvironment{figure}{\let\caption\NAT@figcap}{}    
\newcommand{\NAT@figcap}[2][]{{%
    \refstepcounter{figure}
    \ifthenelse{\value{figure}=1}{
        \noindent%
    }{
    }
    \newline
    \sffamily\noindent\textbf{Figure \arabic{figure}}\hspace{1em}#2}
    \newline
    }
\newcounter{extfig}
\renewenvironment{extfig}{\let\caption\NAT@extfigcap}{}    
\newcommand{\NAT@extfigcap}[2][]{{%
    \refstepcounter{extfig}
    \ifthenelse{\value{extfig}=1}{
        \noindent%
    }{
    }
    \newline
    \sffamily\noindent\textbf{Supplementary Figure \arabic{extfig}.}\hspace{1em}#2}
    \newline
    }
\makeatother
\title{Evidence for the start of planet formation in a young circumstellar disk}

\author{%
	Daniel Harsono$^{1}$\textsuperscript{*}, 
    Per Bjerkeli$^{2,3}$, 
    Matthijs H.~D.~van der Wiel$^{4}$, 
    Jon P.~Ramsey$^{3}$, 
    Luke ~T.~Maud$^{1}$, 
    Lars E.~Kristensen$^{3}$,
    Jes K.~J{\o}rgensen$^{3}$
}
\begin{document}

\maketitle

\begin{affiliations}
 \item Leiden Observatory, Leiden University, P.O. box 9513, 2300 RA Leiden, 
 	The Netherlands
    
 \item  Department of Space, Earth and Environment, Chalmers University of 
 	Technology, Onsala Space Observatory, 43992 Onsala, Sweden 
    
 \item Centre for Star and Planet Formation, Niels Bohr Institute \& Natural History 
 		Museum of Denmark, University of Copenhagen, {\O}ster Voldgade 5--7, 1350 
        Copenhagen K, Denmark
 
 \item ASTRON, the Netherlands Institute for Radio Astronomy, P.O.~Box 2, 7990 AA 
 		Dwingeloo, The Netherlands
 
\end{affiliations}

\begin{abstract} 
The growth of dust grains in protoplanetary disks is a necessary first step towards 
planet formation\citemain{testi14}.
This growth has been inferred via observations of thermal dust 
emission\citemain{beckwith90} towards mature protoplanetary systems (age $\mathbf{>\! 2}$ 
million years) with masses that are, on average, similar to Neptune\citemain{pascucci16}.
In contrast, the majority of confirmed exoplanets are heavier than 
Neptune\citemain{exoplanets.eu}.
{Given that young protoplanetary disks are more massive than their mature counterparts, this} suggests that planet formation starts early, but evidence for grain growth that 
is spatially and temporally coincident with a massive reservoir in young disks remains 
scarce.
Here, we report observations on a lack of emission of carbon monoxide isotopologues 
within the inner $\sim$15 au of a very young (age $\sim$100,000 years) disk around the 
Solar-type protostar TMC1A. 
By using the absence of spatially resolved molecular line emission to infer the 
gas and dust content of the disk, we conclude that shielding by millimetre-size grains 
is responsible for the lack of emission. 
This suggests that grain growth and millimetre-size dust grains can be spatially 
and temporally coincident with a mass reservoir sufficient for 
giant planet formation.  
Hence, planet formation starts during the earliest, embedded phases in the life 
of young stars.
\end{abstract}


TMC1A (IRAS 04365+2535) is a young, low-mass ($\sim 1 M_{\rm Sun}$) 
protostellar system with a 100 au, rotationally  supported disk in the nearby 
Taurus star-forming region\citemain{harsono14,aso15}. 
Based on multi-wavelength continuum observations, TMC1A has been classified as a 
Class I protostar\citemain{lada87}, implying that the protostar and disk system 
is surrounded by a substantial infalling circumstellar envelope from which 
it is still accreting material\citemain{hogerheijde98,aso15}.  
Using the Atacama Large Millimetre/sub-millimetre Array (ALMA), thermal dust 
continuum emission was observed at a wavelength of 1.3 mm along with the $J$ = 2 -- 1 
rotational transition of the carbon monoxide isotopologues $^{13}$CO and C$^{18}$O. 
The disk in TMC1A is driving a wind\citemain{bjerkeli2016} and large-scale 
outflow\citemain{chandler95a,aso15} which are carving a cavity in the envelope. 
Our observations spatially resolve the inner regions of the disk with an 
angular resolution of 0.06$'' \times$ 0.05$''$(PA=$25^{\circ}$), 
corresponding to a spatial scale of 8.4 au $\times$ 7.0 au when adopting a distance 
of 140 parsecs to the Taurus Molecular Cloud.

Previously, we reported resolved observations of a disk wind as revealed by 
$^{12}$CO molecular emission in this system\citemain{bjerkeli2016}.
In this letter, we instead focus on the carbon monoxide isotopologue ($^{13}$CO 
and C$^{18}$O) molecular emission relative to the dust emission in the inner disk 
region ($<50$ au radius).
After re-calibration and inclusion of a previously omitted dataset, we have been 
able to significantly improve the signal-to-noise ratio (see Methods), revealing 
substructure that was only hinted at in the previous study.

Figure 1 shows the 1.33 mm thermal emission from the dust within a 100 au radius 
of the protostar. 
The dust emission peaks strongly within 5 au of the protostar and there is a
shoulder\citemain{bjerkeli2016} at radius $\sim 20$ au, beyond which the dust 
optical depth suddenly decreases ({Supplementary} Figure 1). 
The dust emission is otherwise smooth without any indication of other dips, bumps 
or non-axisymmetric features (such as observed towards, e.g., 
HL Tau\citemain{hltau2014, kzhang15}, TW Hya\citemain{andrews16} or 
Elias 2-27\citemain{lperez16sci}).
From the thermal dust emission, we derive {a lower} limit to the total 
disk mass (dust + gas) of 0.01--0.04 $M_{\rm Sun}$ (10--40 $M_{\rm Jup}$; 
see Methods).

The molecular gas emission is, meanwhile, much more extended than the 
dust continuum (Fig.\ 1); $^{13}$CO and C$^{18}$O emission is visible up to 
a radius of 70 au whereas the mm-dust emission is concentrated inside 40 
au\citemain{bjerkeli2016}. 
The molecular gas emission is, however, only observed (with a spectral 
resolution of 0.35 km s$^{-1}$) down to a radius of $\sim$15 au at velocities 
$> 2$ km s$^{-1}$ relative to the systemic velocity\citemain{harsono14} 
of 6.4 km/s (Fig.\ 1). 
Velocity profile analysis using Keplerian masking of the $^{13}$CO and
C$^{18}$O emission (Fig.\ 2) indicate a 0.4--0.8 $M_{\rm Sun}$ central
protostar\citemain{harsono14,aso15,bjerkeli2016}.

Both $^{13}$CO and C$^{18}$O molecular gas emission are strongly suppressed
inside a 15 au radius. 
If molecular line emission was present in this region, it would be Doppler-shifted 
with respect to the systemic velocity and would, therefore, not be obscured by the 
foreground material. 
Moreover, any gas along the line of sight would have difficulties absorbing the 
emission due to the large velocity and temperature gradient along the line of sight.  
Thus, the suppression must instead be a result of the local disk conditions. 
From the integrated molecular line emission, a gas mass of
0.03--0.07 $M_{\rm Sun}$ (30--70 $M_{\rm Jup}$; see Methods) is derived.
Since the molecular emission is missing in the inner region, this value 
is, in fact, a lower limit to the gas mass if opaque dust is the cause of 
the suppression.

When radiation from a column of material (i.e., gas or dust) is observed at 
a particular wavelength, only emission that originates from 
line-of-sight optical depths $\lesssim$1 is visible to the observer.  
All other radiation at the same wavelength, regardless of source, becomes 
obscured by the preceding material.  
As we are observing both molecular lines and dust (1.33 mm) continuum along 
the line of sight, {if the dust column becomes optically thick (i.e., 
$\tau_{\rm 1.33 mm} \gtrsim 1$), the saturated continuum emission will overwhelm any 
line emission at a similar wavelength, even if it lies closer to the observer.}
The total mass (gas + dust) values presented above have large uncertainties due 
to the assumed gas-to-dust ratio and average temperature. 
As shown by Bjerkeli et al.\citemain{bjerkeli2016}, the $^{12}$CO 
emission at high velocities are detected spatially separated from the 
continuum emission. 
In addition, observed fundamental ro-vibrational\citemain{herczeg11} lines of CO 
point to a hot CO gas reservoir in the inner 0.2 au, implying that 
UV/X-ray radiation originating from stellar accretion cannot be responsible for the
entirely missing $^{13}$CO and C$^{18}$O emission. 
By comparing our observations with radiative transfer 
models\citemain{bruderer12,hyperion}, the suppression of molecular line 
emission in the presence of dust permits us to infer the physical properties
of this young disk and thus understand the relationship between the observed emission 
and the distribution of gas and dust in TMC1A.

Our fiducial model is a parametrised protoplanetary disk with a 
standard dust population and canonical gas-to-dust ratio (see Methods).
From the derived disk mass, our fiducial model predicts readily
observable $^{13}$CO and C$^{18}$O gas emission from the inner $\sim$15 au (Fig.\ 3, 
panel b); this model cannot provide the high optical depth required to suppress 
the molecular emission. 
Models of higher mass disks meanwhile suggest that a total disk mass of 
$> 0.2 M_{\rm Sun}$ ($\sim 200 M_{\rm Jup}$; as compared to a 0.4--0.8 
$M_{\rm Sun}$ protostar) is needed to reach the required optical depths to 
suppress the molecular emission (Fig.\ 3). 
Massive disks, however, are gravitationally unstable when the disk-to-stellar 
mass ratio\citemain{lr04} $> 10\%$. 
In this case, the TMC1A disk should then be unstable and display 
non-axisymmetric structures\citemain{lperez16sci} on time scales of decades 
(see Methods), but neither the dust nor gas emission show evidence of such 
features.
If non-axisymmetric structures are present in this disk, they are at lower 
amplitudes than probed by these observations.

A more plausible explanation for the large optical depth is that the dust 
population is not standard and large grains ($>1$ mm) are already 
present, at the very least, inside 30 au.  
Typically, dust mass absorption coefficients due to grains with icy mantles 
and sizes $\leq$ 0.25 microns\citemain{OH94} are sufficient to explain 
sub-millimetre thermal dust emission from protostellar envelopes and 
dense cores\citemain{henning95,launhardt13}, whereas larger grains are 
commonly used to reproduce dust emission from more evolved, Class II 
disks\citemain{ricci10,andrews13}. 
For TMC1A, larger dust grains in the inner regions increase the mass 
absorption coefficient\citemain{andrews09} and our modelling demonstrates that 
this can account for the suppression of molecular emission (Fig.\ 3) while 
also providing a lower disk-to-stellar mass ratio ($\sim 5\%$; see Methods).  
The suppression in molecular emission can also be reproduced if we adopt an 
unusually flat (i.e.\ a reduced vertical scale height; see Methods) and 
cold disk model.  
The properties of cold and flat disks however provide a better match 
to less massive disks around mature Class II sources, and are 
inconsistent with the thermal structure of such young disks\citemain{visser09},
especially that of TMC1A, which is known to still be accreting from 
the surrounding envelope\citemain{aso15}.

Following all of these considerations, we conclude that the presence of grains 
$\geq$ mm-size are the cause of the suppression of the molecular gas emission. 
Our spatially resolved data demonstrates the effect of large grains on 
the emergence of molecular gas emission on Solar System scales ($< 30$ au) in 
this young disk. 
In contrast to older Class II disks\citemain{pascucci16}, we find that the young 
protoplanetary disk around TMC1A has sufficient mass to form multiple 
Jupiter mass planets.  
The similarities in structure (e.g.\ scale height of the gas disk, 
radial exponential tail, surface density power-law index) and dust composition 
(small and large grains populations) inferred for this particular young, 
embedded protostellar disk, relative to more mature Class II disks, meanwhile implies
that the seeds for planet formation are already present in the early stages 
of protoplanetary disk evolution.

Millimetre-size grains in such a young disk is of particular interest for planet 
formation. 
Core accretion models of giant planet formation rely on the availability of a 
large number of planetesimals\citemain{mizuno80} to explain the growth of planets.  
Astrophysical pebbles are, meanwhile, typically between mm- to cm-size\citemain{lambrechts14}.
Pebbles accrete onto planetary bodies more efficiently than planetesimals owing to
their much larger numbers and partial coupling to the gas, resulting in planetary
growth time scales well within the disk lifetime\citemain{lambrechts12,ida16}.  
Indeed, the presence of a significant population of mm-size dust in a $\sim\! 10^5$ 
year old disk is in agreement with time scales from recent models of dust 
growth\citemain{testi14}.  
It is also consistent with the measured size distribution of the thermally processed 
dust grains found in chondritic meteorites known as chondrules\citemain{davis14} 
and, indeed, recent age estimates for the oldest known examples\citemain{connelly12}. 
{
Suppression of molecular emission by dust provides an independent means 
to investigate the presence of grain growth in spatially resolved studies 
of young disks. Therefore, o}bservations towards other protostellar systems
similar to the one we presented here 
should be undertaken in order to determine how common mm-size dust is in the 
inner regions of very young disks, {thus} leading to a better understanding 
of when, and how frequently, young disks are forming planets.
\newpage
%
%
\begin{figure}
\centering
\begin{tabular}{c}
\includegraphics[width=0.8\textwidth]{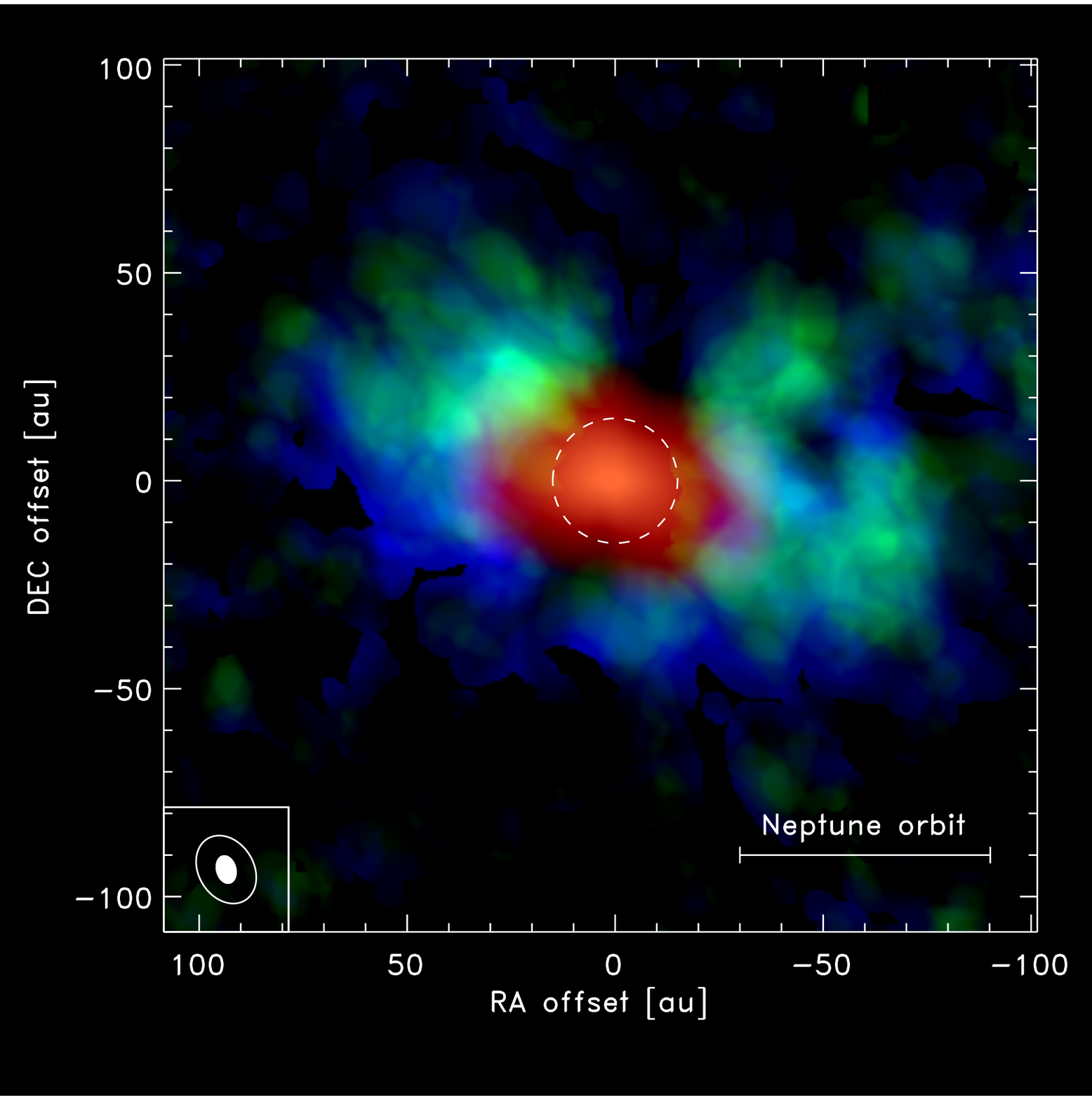}
\end{tabular}
\caption{
\textbf{
Dust continuum and integrated $^{13}$CO and C$^{18}$O emission map of TMC1A.}
The colours represent 1.3 mm dust continuum (red), $^{13}$CO (green) and 
C$^{18}$O (blue) zeroth moment maps.  
The intensities are scaled linearly with the dust continuum/integrated line fluxes. 
The peaks of the maps are 13.9 mJy beam$^{-1}$, 33.8 mJy beam$^{-1}$ km s$^{-1}$, 
and 17.8 mJy beam$^{-1}$ km s$^{-1}$ for the dust continuum, $^{13}$CO, and 
C$^{18}$O zeroth moment maps, respectively.  
The molecular emission falls off steeply at radii $\lesssim$30 au, and 
is completely absent in the inner 15 au as indicated by the dashed line contour. 
The diameter of Neptune's orbit (60 au) is indicated at the bottom right. 
The synthesized beam sizes, shown in the bottom left, are $0.06''\times 0.05''$
(position angle of 25$^{\circ}$) for the dust continuum and 
$0.12''\times 0.095''$ (position angle of 31$^\circ$) for the molecular lines.
}
\label{fig:Fig1}
\end{figure}
\begin{figure}
\centering
\begin{tabular}{c}
\includegraphics[width=0.5\textwidth]{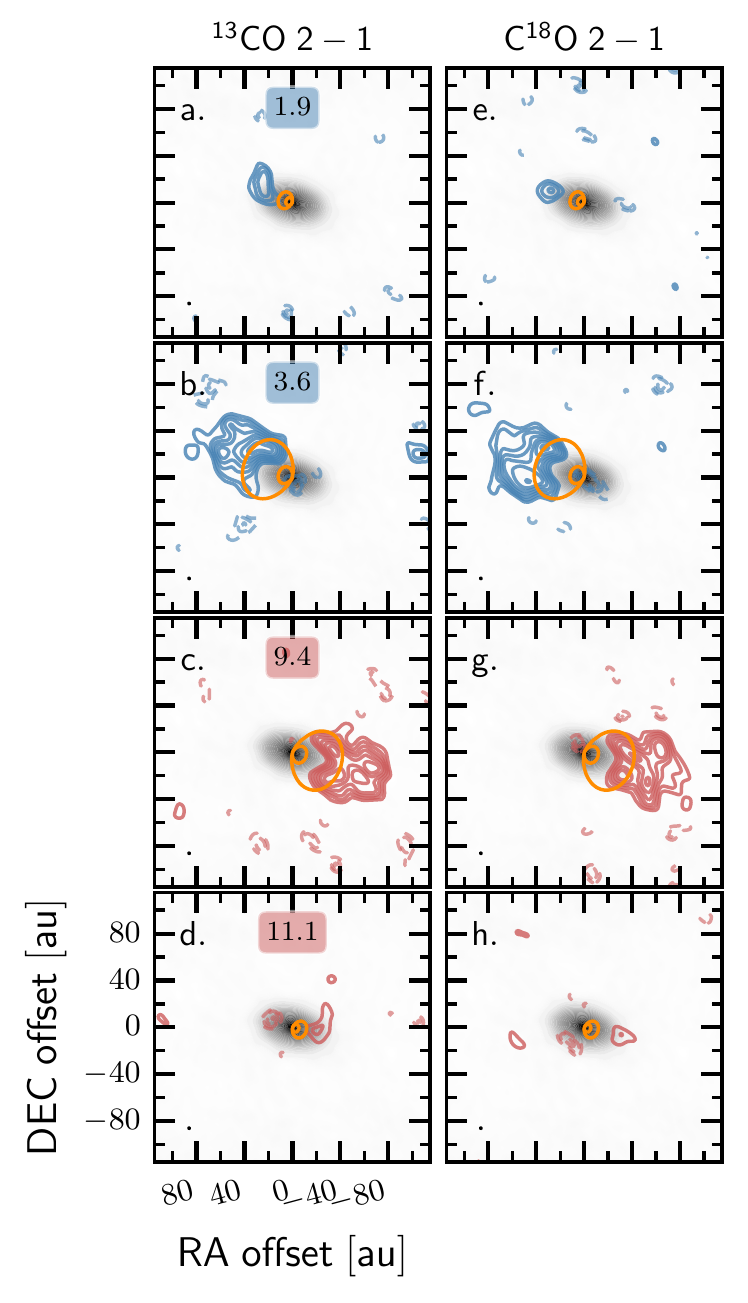}
\end{tabular}
\caption{
\textbf{$^{13}$CO and C$^{18}$O channel maps.} 
Line contours show integrated emission from $3\sigma$ to $15 \sigma$ at every 
$\sigma$ for the $^{13}$CO ($\sigma$ = 2.2 mJy beam$^{-1}$ km s$^{-1}$; 
panels a - d) and C$^{18}$O channel maps ($\sigma$ = 1.6 mJy beam$^{-1}$ km s$^{-1}$; 
panels e - h). 
Integrated maps begin at 1.0 km s$^{-1}$ to 4.5 km s$^{-1}$ for the blue 
side and 8.5 km s$^{-1}$ to 12 km s$^{-1}$ for the red side.
Each panel shows the emission integrated over 5 channels (one channel is 
0.35 km s$^{-1}$).  
Absorption is denoted with dashed contours starting from -5$\sigma$ 
up to -2$\sigma$ in steps of 0.5$\sigma$.
Dust continuum is shown in gray from 5$\sigma$ up to the
maximum peak intensity. 
The midpoint velocity of each channel is shown at the top of each panel in 
the left column.
Keplerian iso-velocity contours for a disk radius of 200 au and a 
$0.5 M_{\rm Sun}$ central source inclined at 51$^{\circ}$ with 
respect to the plane of the sky are shown in orange. 
}
\label{fig:Fig2}
\end{figure}
\begin{figure}
\centering
\begin{tabular}{c}
\includegraphics[width=1.0\textwidth]{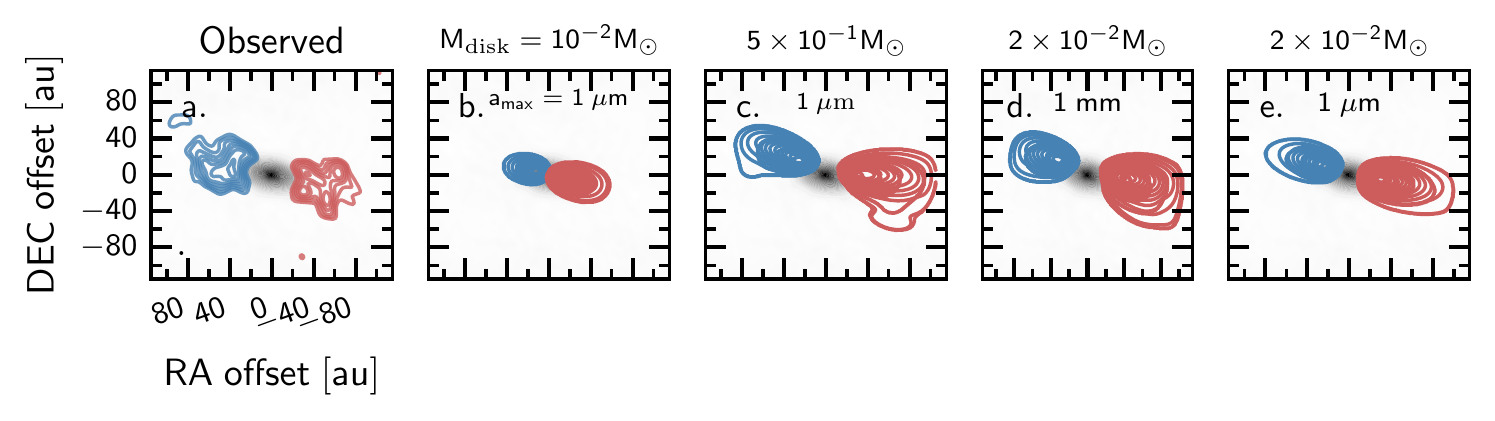}
\end{tabular}
\caption{
\textbf{Observed and simulated C$^{18}$O emission maps.} 
The molecular emission is integrated between 2 to 4.5 km s$^{-1}$ (blue) and 
8 to 10.5 km s$^{-1}$ (red).
The observed dust continuum is shown in gray with a linear scaling from 
5$\sigma$ to the maximum.  
The observations are shown in panel (a). 
Four radiative transfer models (b - e) with different 
parameters are shown: b) A disk with mass $0.01 M_{\rm Sun}$; note that both
red- and blue-shifted emission is expected to extend down to the central protostar.
The depression in the molecular emission can be reproduced by 
c) a massive disk of $0.5 \ M_{\rm Sun}$, d) a population of large grains 
($a_\mathrm{max} = 1$ mm) or e) a flat, cold disk with parameters similar to
more mature Class II sources.  
The adopted disk masses and maximum grain sizes are indicated in each panel.  
With the exception of panel (d), the models utilise small, bare dust grains 
($a_{\rm max} = 1 \ {\rm \mu m}$) (see Methods).
}
\label{fig:Fig3}
\end{figure}

\newpage
\begin{extfig}
\centering
\begin{tabular}{c}
\includegraphics[width=0.5\textwidth]{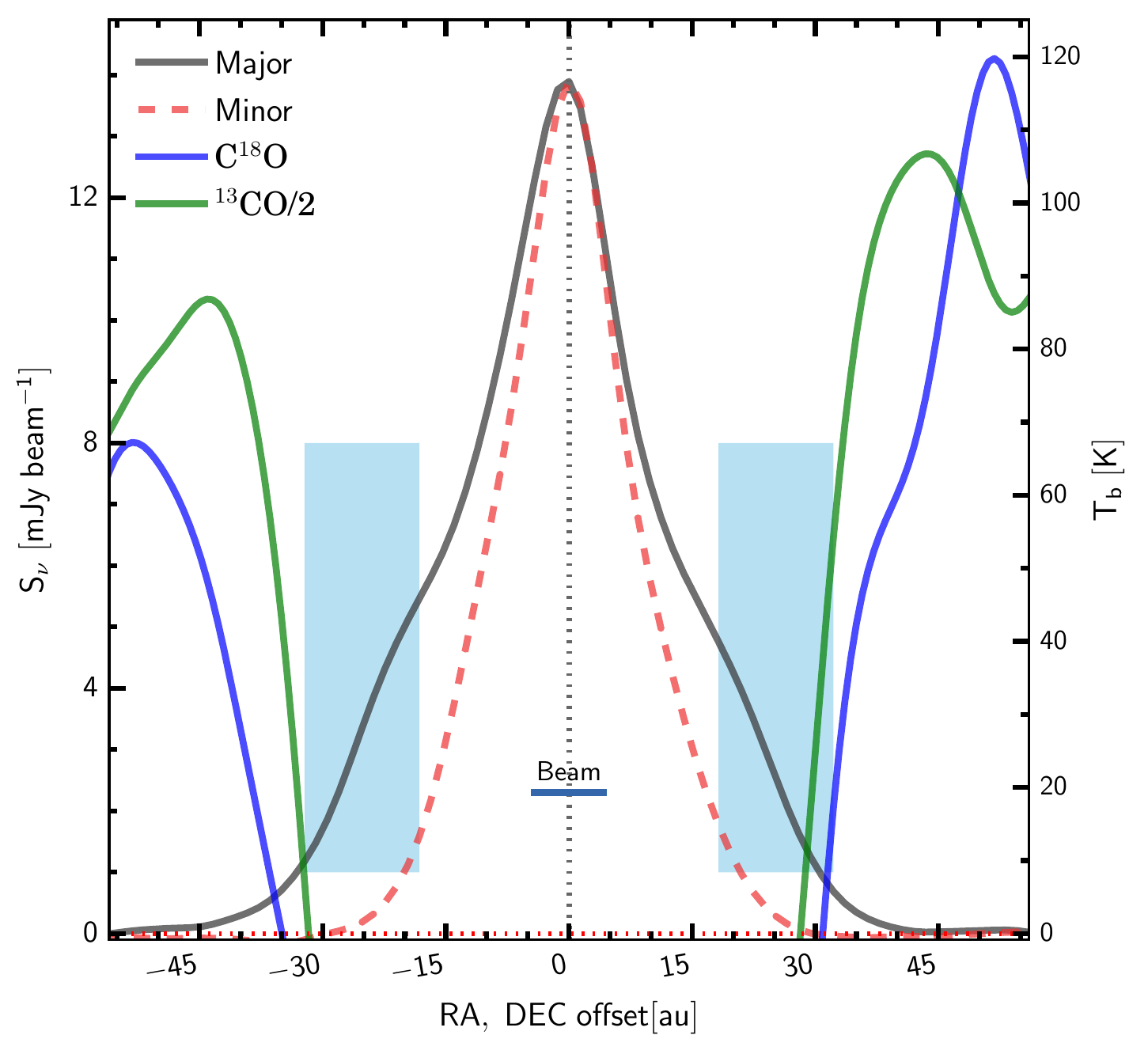}
\end{tabular}
\caption{
\textbf{Intensity profile of the dust continuum.} 
The brightness profiles along major (PA = 76$^{\circ}$ E of N; black) 
and minor ( PA = 166$^{\circ}$ E of N; red) axes of the disk are shown; 
the shoulder is located at $\sim 20$ au.  
The size of the beam is indicated at the bottom of the figure. 
The corresponding brightness temperature of the continuum is shown on the 
right axis.
The blue and green lines show brightness profiles for the zeroth moment maps 
of C$^{18}$O and $^{13}$CO scaled by 2, respectively, along the major axis.  
Two blue-shaded regions are shown to illustrate the approximate locations of 
the shoulder and the start of the gas emission hole.
}
\label{fig:extFig2}
\end{extfig}
\begin{extfig}
\centering
\begin{tabular}{c}
\includegraphics[width=0.8\textwidth]{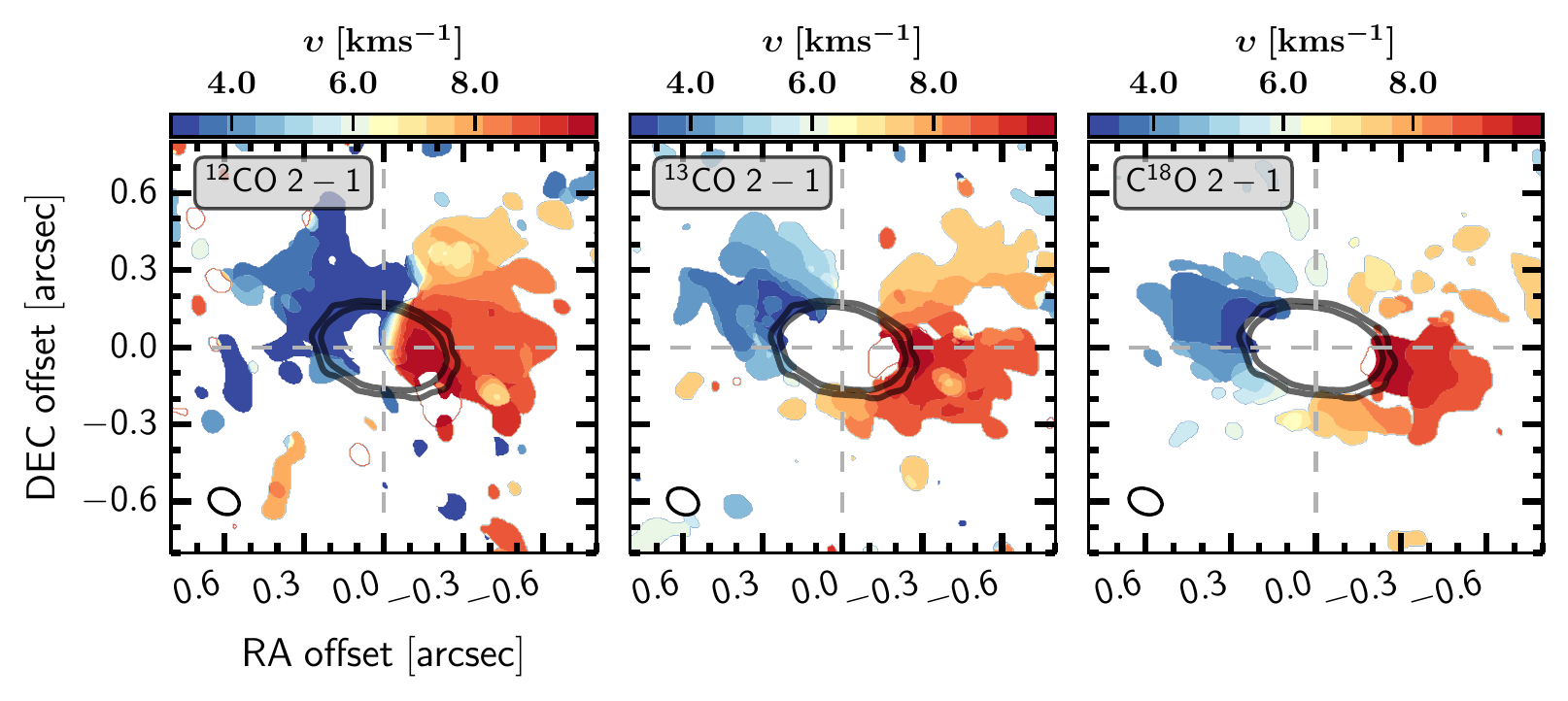}
\end{tabular}
\caption{
\textbf{First moment maps of the CO isotopologues.} 
The blue- and red-shifted emission of the C$^{18}$O (middle) and 
$^{13}$CO (right) molecular emission show a clear rotating structure outside of 
the dust continuum, denoted with black contours.
These maps were constructed only taking into account molecular emission 
from 2 to 11 km s$^{-1}$.
The $^{12}$CO (left) first moment map shows rotating gas shifted north with 
respect to the C$^{18}$O map. 
It also contains a highly blue-shifted gas component that is not seen in 
$^{13}$CO or C$^{18}$O.
}
\label{fig:exFig3}
\end{extfig}
\begin{addendum}	
	\item This paper makes use of the following ALMA data: ADS/JAO.ALMA\#2015.1.01415.S. 
	ALMA is a partnership of ESO (representing its member states), NSF (USA) and 
	NINS (Japan), together with NRC (Canada) and NSC and ASIAA (Taiwan) and KASI 
	(Republic of Korea), in cooperation with the Republic of Chile. The Joint 
    ALMA Observatory is operated by ESO, AUI/NRAO and NAOJ. The authors would 
	like to thank Allegro, the European ALMA Regional Centre node in the Netherlands, 
    for providing the facilities and assistance in re-calibrating and imaging of the 
    data. 
    We thank the anonymous referees for their helpful comments that have greatly 
    improved the manuscript.
    Furthermore, DH thanks Michiel Hogerheijde, Arthur Bosman and Ewine van Dishoeck 
    for interesting discussions. 
    DH is supported by EU ERC Advanced Grant 291141 ``CHEMPLAN'' and by a 
    KNAW professor prize awarded to E.\ van Dishoeck.
    DH and LTM are part of Allegro, which is funded by the Netherlands Organisation for 
    Scientific Research (NWO). 
    PB acknowledge the support by the Swedish Research Council (VR) 
    through contracts 2013-00472 and 2017-04924. 
    The group of JKJ acknowledges support from the European Research Council (ERC) 
    under the European Union's Horizon 2020 research and innovation programme 
    (grant agreement No~646908) through ERC Consolidator Grant ``S4F''. Research 
    at the Centre for Star and Planet Formation is funded by the Danish National 
    Research Foundation (DNRF97).

    \item[Author Contributions] DH and LTM were responsible for the data 
    re-calibration and reduction. 
    DH was responsible for the analysis and wrote the manuscript together with 
    PB, MHDvdW and JPR. MHDvdW and PB composed the observing proposal, 
    with contributions from DH, JPR and JKJ.  
    All authors contributed at various stages to the data analysis, 
    discussed the results and contributed to the manuscript.

 	\item[Author Information] Reprints and permissions information is available 
    at www.nature.com/reprints. The authors declare that they have no competing 
    financial interests. Correspondence and requests for materials should be addressed 
    to D.\ Harsono (email: harsono@strw.leidenuniv.nl).

    { 
    \item[Competing Interests] The authors declare no competing interests.
    }
    
\end{addendum}
\bibliographystylemain{naturemag}
\bibliographymain{biblio.bib}

\begin{methods}

\paragraph{Observations and calibration.}  

TMC1A was observed with ALMA on three different dates: October 16 2015 
(Execution Block 1; EB1), October 23 2015 (EB2) and October 30 2015 (EB3). 
The last two execution blocks were taken under excellent weather conditions 
(water vapour $<$ 1 mm). 
Bjerkeli et al.\citemain{bjerkeli2016} presented only execution blocks 
EB2 and EB3 with lower water vapour and relatively good stability.  
Three SPWs were dedicated to observe the $^{12}$CO 2-1 (230.538 GHz, 
$E_{\rm up}$ = 16.60 K), $^{13}$CO 2-1 (220.399 GHz, $E_{\rm up}$ = 15.87 K), 
and C$^{18}$O 2-1 (219.560 GHz, $E_{\rm up}$ = 15.81 K) transitions.  
As poor conditions can lead to defects in images\citemethods{clcarilli99}, 
the phase centre of TMC1A was derived from the EB3 data, which had the best 
atmospheric stability.

The entire data set (all three EBs) were re-calibrated with CASA v4.7.2.  
Instead of using the observed integration time (3s) to obtain the phase solutions, 
we opted to use 12s in order to obtain better solutions with higher 
signal-to-noise.
The flux calibrators were manually checked for consistency between execution 
blocks.
Furthermore, we have used a non-standard water-vapour radiometer (WVR) 
scaling\citemethods{maud17wvr} to improve the phase delays for each antenna due 
to atmospheric water content.
Scaling of the WVR solutions is the only way to improve the phase stability on 
very short timescales (integration time) which is typically not possible except 
for very strong targets\citemethods{maud17wvr}.  
We refer to Bjerkeli et al.\citemain{bjerkeli2016} for imaging details.

Since the 1.3 mm continuum emission of TMC1A is detected with a high image 
signal-to-noise ($S/N > 100$), it is possible to calibrate the data set against 
itself (``self-calibration'').
The improvements introduced by WVR corrections resulted in higher signal-to-noise 
data and images, allowing us to perform self-calibration on the target down to a 20s 
timescale.
At first, large time scale phase corrections (5 minutes) were applied to all three 
data sets in order to minimize the long period atmospheric phase decoherence. 
After the final self-calibration, the phase de-coherence that usually spreads 
flux around an image with increasing noise\citemethods{clcarilli99} was reduced. 
The resulting noise levels amount to an improvement of a factor four and 
two relative to the previously presented continuum image and 
CO maps\citemain{bjerkeli2016}, respectively.

\paragraph{Thermal dust emission.}

The dust continuum emission provides a view into the density structure of the disk. 
After calibration and analysis, the resulting synthesized beam ($0.06'' \times 0.05''$, \\
PA = $25^{\circ}$) is sufficient to spatially resolve the full extent of 
the dust disk ($\sim 0.3''$ radius). 
Due to the small, synthesized beam, a common fixed phase centre was introduced 
to minimize imaging artefacts due to phase centre offsets between execution blocks.
The peak of the continuum intensity is 13.9 mJy beam$^{-1}$ (Figure 1) with an 
integrated flux density of 145$\pm 2$ mJy at 225 GHz inside a $1''$ box. 
The image noise level is 45 $\mu$Jy beam$^{-1}$. 
The disk is inclined at 51$^{\circ}$ with respect to the plane of the sky 
(0$^{\circ}$ is face-on)\citemain{harsono14,aso15,bjerkeli2016}, 
as derived from the ratio of the major to minor axis (deconvolved 
$FWHM = 0.24'' \times 0.15''$) of the dust continuum image. 
The PA was determined from a 2D Gaussian fit to the emission. 
{Supplementary} Figure 1 shows the brightness profile along the major 
(PA = 76$^{\circ}$) and minor (PA = 166$^{\circ}$) axes; 
it is symmetric up to $\sim$20 au, where a shoulder is visible 
due to a change in the optical depth.

The dust continuum is observed at 225 GHz (1.33 mm), and may be contaminated 
by free-free emission from unresolved accretion shocks at the stellar surface. 
However, Scaife et al{.}\citemethods{scaife12} have shown that the 
mm-continuum emission from Class I sources similar to TMC1A are instead dominated 
by the dust thermal emission\citemethods{ubach17}. 
By comparing the images obtained from applying different weightings (superuniform versus 
Briggs weighting\citemethods{briggs95} of 1), at most, a flux of 3 mJy can be 
attributed to free-free emission, which is negligible compared to the total flux of 
the disk. 
Therefore, the total disk mass (gas + dust) can be safely estimated from the 
dust continuum by assuming an average dust temperature and total mass absorption 
coefficient $\kappa_{\nu}$. 
Under the assumption of optically thin dust thermal emission, the total
mass is given by: 
\begin{equation}
  M_{\rm total = gas + dust} = \frac{S_{\nu} d^2}{\kappa_{\nu} B_{\nu} 
  \left ( T_{\rm d} \right )},
  \label{eq:estimate_mass}
\end{equation}
where $S_{\nu}$ is the flux density in Janskys, $\kappa_{\nu}$ is the mass 
absorption coefficient at frequency $\nu$ corrected for a standard 
gas-to-dust mass ratio\citemethods{draine07} (100) and $B_{\nu}$ is 
the Planck function at the dust temperature $T_{\rm d}$ assuming the gas temperature 
is coupled to the dust.

One of the quantities typically used to estimate the evolutionary stage of 
a protostellar system is the disk mass\citemethods{hueso05, robitaille06, takakuwa12}.
A total disk mass (gas$+$dust) of 0.012--0.036 $M_{\rm Sun}$ is derived using dust 
temperatures in the 30 to 80 K range as indicated by the measured 
brightness temperature \\ ($T_{\rm B} = \frac{\lambda^2}{2 k_{\rm B}} I_{\nu}$ 
= 30--120 K in the inner 30 au).
The adopted total mass absorption coefficient\citemain{beckwith90} is 
$\kappa_{\nu}$ = 0.01 ($\nu$/230GHz) cm$^{2}$ g$^{-1}$, which is similar to 
a $\kappa_{\nu}$ value of 0.0085 cm$^{2}$ g$^{-1}$ at 1.3 mm from 
Ossenkopf \& Henning\citemain{OH94}.
For mature Class II disks, a mass absorption coefficient of $\sim$0.02 cm$^{2}$ 
g$^{-1}$ is typically used\citemain{ricci10}$^{,}$\citemethods{andrews11}.
By adopting dust opacities appropriate for Class II disks, a factor of two lower 
disk mass is obtained for TMC1A.

Recent studies of the unresolved dust continuum and multi-wavelength spectral index 
have found evidence for grain growth in Class 0 and I 
objects\citemethods{cflee08,prosac09,kwon09,scaife12,chiang12,tobin13,miotello14}.  
Indeed, Lee et al.\citemethods{cflee17} recently found evidence for an optically thick
dust continuum in the edge-on protostellar system HH 212. 
However, the observation of emission from the surface layers of a disk in such a 
configuration is not trivial. 
In part due to its lower inclination, the surface layers of the TMC1A disk are 
directly observable and permit the observation of the optically thick dust continuum 
and the subsequent effect on the emergence of molecular emission.

\paragraph{Dust mass absorption coefficients and properties.}

We explored the effects of various dust mass absorption coefficients (opacities) 
on the emergence of molecular emission.
They were calculated using the Bohren and Huffman Mie (``BHMie'') 
code\citemethods{bhmie83} distributed with {\it Hyperion}\citemain{hyperion}. 
For small grains, the sizes are between 0.01 -- 1 microns following Andrews 
et al.\citemethods{andrews11}. 
For large grain opacities, we vary the maximum grain size, $a_{\rm max}$, between 
1 mm and 10 cm. 
Dust opacities with $a_{\rm max} = 10$ cm lead to a lower dust mass absorption 
coefficient at millimetre wavelengths.
The opacities presented here are normalised to the dust mass.
The limited wavelength coverage of our data set can only constrain the maximum grain size 
to a few millimeters. 
Thus, we adopt $a_{\rm max}$ of 1 mm for simplicity. 
Optical constants were taken from Weingartner \& Draine\citemethods{wd01} 
and Draine\citemethods{draine03}.

We have compared our calculated opacities with Ossenkopf \& Henning 
opacities\citemain{OH94} with a standard\citemethods{mrn77} size distribution and 
a thin ice mantle at densities of $10^{6}$ cm$^{-3}$. 
The mass absorption coefficient differs by a factor of two (0.899 cm$^{2}$ 
g$^{-1}$ with ice mantle versus $\sim$2 cm$^{2}$ g$^{-1}$ without) at 1.33 mm. 
While Ossenkopf \& Henning opacities describe the dust properties of 
protostellar objects, BHMie calculations are suitable for T-Tauri 
disks\citemain{beckwith90}$^{,}$\citemethods{draine06}.  
Since the presence of ice lowers the mass absorption coefficient, in order to 
maximize the absorption coefficient, only bare grains are considered.
The use of bare grains furthermore permit the derived disk mass to be minimized. 
Despite these considerations, Ossenkopf \& Henning opacities alone 
cannot provide the larger maximum grain size required to reach the derived 
optical depths.

\paragraph{Molecular emission.}

Figure 1 (blue and green colours) shows the zeroth moment maps for the 
observed CO isotopologue lines. 
Moment maps were constructed by considering pixels whose intensities 
$>3\sigma$ ($1 \sigma$ = 2.1, 2.3, and 1.7 mJy beam$^{-1}$ per 0.35 km s$^{-1}$ 
channel for $^{12}$CO, $^{13}$CO, and C$^{18}$O, respectively) and 
velocity channels between 2 - 11 km s$^{-1}$. 
For comparison with the known wind component\citemain{bjerkeli2016}, which is 
best traced by the $^{12}$CO 2--1 line, the line profiles of the three CO 
isotopologs were extracted from a 2$''$ box (from -1$''$ to 1$''$).  
$^{12}$CO molecular emission is detected between -3.8 -- 14.4 km s$^{-1}$. 
On the other hand, the C$^{18}$O 2--1 line is detected between
2.15 -- 10.55 km s$^{-1}$ and traces the quiescent gas of the Keplerian disk.
The characteristic butterfly pattern of a Keplerian velocity profile is clearly
visible in the channel maps (Figure 2 and {Supplementary} Figure 2). 
Both zeroth (Figure 1) and first ({Supplementary} Figure 2) moment maps show 
that $^{13}$CO and C$^{18}$O emission is absent within the inner 15 au.

The integrated line fluxes for $^{13}$CO and C$^{18}$O lines are $0.99 \pm 0.2$ 
Jy km s$^{-1}$ and $1.04 \pm 0.1$ Jy km s$^{-1}$, respectively, within a $1''$ box. 
The 1$\sigma$ error was determined from the sum of an averaged root mean square
(r.m.s.) over the spectral range including a 20\% flux error. 
In comparison to the observations of Harsono et al.\citemain{harsono14}, the similarity 
in the $^{13}$CO and C$^{18}$O fluxes is due to the missing short spacing.  
It was already hinted at that the C$^{18}$O emission is more compact than the 
$^{13}$CO emission; thus, $^{13}$CO is affected by the resolved-out large 
structure\citemain{harsono14} more strongly than the C$^{18}$O.
Assuming optically thin emission, adopting a CO abundance of 10$^{-4}$ with 
respect to H$_2$, a $^{16}$O/$^{18}$O isotopic ratio\citemethods{wilson94} of 560 
and a rotational temperature between 40 to 80 K\citemethods{harsono13}, the
C$^{18}$O line gives a derived gas mass of $0.03-0.07 M_{\rm Sun}$. 
These values are roughly comparable to the gas mass derived from the dust 
continuum flux assuming a gas to dust ratio of 100.  
We note that a lower CO abundance would result in an even higher disk mass.

To obtain the disk structure, we have fitted the continuum image using Least 
Square Minimization\citemethods{scipy} and a Markov Chain Monte Carlo package 
(``{\it emcee}''\citemethods{emcee}).
The minimization is performed on pixels whose intensities $\gtrsim$5$\sigma$. 
The intensity profile is fitted with a power-law disk including 
an exponential taper:
\begin{eqnarray}
\label{eq:sigmar}
\Sigma \left ( R \right ) & = & \Sigma_0 \left ( \frac{R}{R_0} \right )^{-p} 
    \exp \left ( - \left(\frac{R}{R_0}\right)^{2 - p} \right ); \\
\label{eq:tempr}
T \left ( R \right ) & = & T_0 R^{-q}.
\end{eqnarray}
We find that a disk characterized by $p\sim 0.6^{+0.1}_{-0.1}$ and 
$q\sim 0.6^{+0.1}_{-0.1}$ can reproduce the continuum intensity profile.  
We also tried a simple two-grain population model: 
\begin{eqnarray}
\Sigma_{\rm tot} = \Sigma_{\rm small} + \Sigma_{\rm large},
\end{eqnarray}
where each $\Sigma$ is described by Equation \ref{eq:sigmar}. 
This returned a flat $p \lesssim 0$ as the best-fitting model. 
{
For completeness, we also tried disk models without a radial taper.}
{Such models favour a steep radial power-law index, $p\sim 5$, in order to obtain a 
steep mass distribution that achieves the necessary high optical depth at the 
shoulder ($\sim$15 au).}
Guided by the above parameters, we also generated a flat, thin disk 
model following Murillo et al.\citemethods{murillo13b} and a density profile
that follows Equation \ref{eq:sigmar}.
These models prefer $p\sim 0.7^{+0.4}_{-0.3}$ and $q \sim 0.9^{+0.1}_{-0.2}$ 
given an opacity described by a combined population of small and large grains.  
Visibility fitting\citemain{harsono14} using power-law disk structures with different 
values of $p$ provided disk parameters that are consistent with the above results.

Our image fitting does not fully explore all of the possible parameter space. 
Instead, it provides guidance for 2D self-consistent radiative transfer models
that we used to understand the emergence of the molecular emission\citemain{bruderer12}.
For simplicity, we chose a power-law index of $p = 1$ scaled by the mass derived from the
observations; a value of $p = 0.5$, meanwhile, produces an intensity profile that 
is too shallow.
The disk gas scale height, $H$, is set to 0.12 au at 1 au. 
Based on the dust, hydrostatic equilibrium for the TMC1A disk predicts a scale height of 
0.03 au.  
However, we found that the gas and the small grains could be characterized with 
a different scale height as compared to a mm-sized dust grain distribution.  
Our goal with the radiative transfer modelling is not to provide a best-fit model 
to the data, but instead to demonstrate how the different parameters affect the 
emergence of CO emission.
The Monte Carlo frequency-dependent radiative transfer code, 
\textit{Hyperion}\citemain{hyperion}, was used to derive the temperature structure 
for the various dust opacities described above. 
A stellar luminosity of 2.7 $L_{\rm Sun}$ and an effective stellar temperature 
of 4000 K were adopted\citemethods{white04}. 
The gas temperature is assumed to be perfectly coupled to the dust temperature. 
{Self-consistent gas temperature calculations lead to a warm molecular layer} 
and will result in a stronger molecular emission, however, we do not observe this 
effect.
There is evidence that points towards lower CO abundances in Class II 
disks\citemain{bruderer12}$^,$\citemethods{ansdell16,kama16,guilloteau16,miotello17}, 
but not necessarily toward younger, Class I disks (van't Hoff et al. in press, 
de la Villarmois et al. in press).   
An escape probability molecular excitation code\citemain{bruderer12} with 
adaptive gridding (to resolve the physical and velocity structure of the disk)
was used to simulate the CO line emission.  
Collisional rate coefficients and Einstein $A$ values were obtained from the Leiden 
Molecular Database\citemethods{lamda,corates}, extrapolated\citemethods{neufeld12} 
to $J$ = 60 to account for the high gas temperatures in the inner few au.  
The partition function is self-consistently calculated by taking the populated 
$J$ levels into account.

Our fiducial disk model has a mass of $10^{-2}$ $M_{\rm Sun}$ and uses small grain 
dust opacities \\ (Fig.\ 3b). 
This model predicts observable red- and blue-shifted molecular emission in the 
inner 15 au, in contrast with the observations.
We intentionally avoid comparing the molecular emission close to the systemic 
velocity since the observations are highly 
dependent on the {baseline coverage} of the data set.
Via additional models, we explored the effects of the disk mass, disk flaring 
angle and dust properties.

We find that the suppression of the molecular emission can be reproduced if the
disk mass $> 2 \times 10^{-1} M_{\rm Sun}$ (Fig.\ 3c). 
More specifically, the optical depth of the C$^{18}$O line can easily be as high 
as 10, which results in a gas mass of $\sim2 \times 10^{-1}$ $M_{\rm Sun}$. 
A low mass disk ($\sim 10^{-2} \ M_{\rm Sun}$) with large grain opacities, 
but otherwise standard parameters, also reproduces the suppression in 
molecular emission (Fig.\ 3d).
As the mass estimates presented earlier were calculated under the assumption 
of optical thinness, this implies that these masses should be considered 
approximate.

A thick, flared disk model (scale height $H$ of 0.3 au at 1 au) predicts 
observable molecular emission inside 15 au radius. 
In a such a flared disk, the location of the gas emission is always above the 
dust photosphere (optical depth $\sim 1$; where the emission originates). 
In the optically thick region of the disk, the peak of the observed emission is 
related to the gas excitation temperature and the temperature at the dust 
photosphere:
\begin{equation}
T_{\rm observed} = T_{\rm excitation} - T_{\rm dust},
\label{eq:t_observed}
\end{equation}
where the intensity is given in terms of temperature since it is thermalised
($S = B_{\nu} \left( T \right)$) and at the Rayleigh-Jeans limit. 
If the two temperatures are similar, the continuum-subtracted gas emission is 
directly related to the difference in kinetic temperature. 
The peak of the gas emission is thus given by $S_{\nu} = B_{\nu} \left 
(T_{\rm excitation} - T_{\rm dust} \right )$, which results in 
negligible molecular emission. 
{
The typical noise level per beam translates to $\gtrsim$4 K uncertainties 
in the temperature, which mean that our data is sensitive to a 4 K difference 
between the dust and gas emitting layers}. 
A flat disk ($H/R$ = 0.06 Fig.\ 3e), meanwhile, produces an absence of emission 
inside 15 au since the gas and dust are vertically confined to a small region 
with a small temperature difference between the two emitting surfaces/photospheres. 
However, such a flat disk is typically found in lighter, mature Class II 
disks\citemain{andrews09} and not in massive, young disks such as TMC1A.

A temperature inversion due to strong viscous heating in the disk could explain 
the absence of molecular emission. 
In accreting disk models\citemethods{harsono15b}, the inversion is typically found 
in regions of the disk $>$300 K.
To obtain a temperature inversion at a distance of 15 au in TMC1A, a very high 
stellar accretion rate ($> 10^{-5} M_{\rm Sun}$ yr$^{-1}$) is required. 
This yields a total luminosity $>$10 $L_{\rm Sun}$, which is inconsistent with 
the measured bolometric luminosity\citemethods{kristensen12} of 2.7 $L_{\odot}$. 
Furthermore, modelling of the water lines\citemethods{yvart16} observed by 
{\it Herschel} {\it HIFI}\citemethods{WISH,kristensen12} also suggests accretion rates 
of 10$^{-6} \ M_{\odot} \ \mathrm{yr}^{-1}$. 
Therefore, a temperature inversion is unlikely to be responsible for 
the missing molecular lines.

\subsection{Dynamical mass of the young star and disk stability.}

The dynamical mass of a protostellar object places important constraints on 
the evolution and structure of the associated protoplanetary disk.
The disk-to-stellar mass ratio also dictates the gravitational stability of the 
disk\citemain{lr04}$^{,}$\citemethods{kratterlodato16}. 
Figure 2 shows the Keplerian iso-velocity contours against the $^{13}$CO 
and C$^{18}$O emission within the same velocity range.
{Supplementary} Figure 2 meanwhile shows the first moment map indicating the 
direction of the Keplerian rotation. 
The iso-velocity contours and subsequent Keplerian velocity 
profile\citemain{bjerkeli2016} indicates that the protostellar mass is between 
0.4--0.8 $M_{\rm Sun}$.
The range in mass results from the consideration that the line emission 
is not optically thin and, since the highly Doppler-shifted emission near 
the protostar is absent, we are limited from placing better constraints on 
the protostellar mass.

Using the disk masses derived from both gas and dust emission, the disk-to-stellar 
mass ratio ranges from 0.02 to 0.04. 
As noted above, the disk mass must be higher ($> 0.2 M_{\rm Sun}$) in order 
to suppress the molecular emission if a standard grain population is used.  
Such a high disk mass would imply that the disk-to-stellar mass ratio is between 
0.25 to 0.625.  
Numerical simulations\citemethods{lr05,cossins09} show that such a high 
disk-to-stellar mass ratio is gravitationally unstable disk and leads to 
non-axisymmetric structures.
The $\alpha$-viscosity\citemethods{ss73} of such a disk is of the order 
0.05 to 0.1, while Kratter et al.\citemethods{kratter10} report $\alpha$ values 
between 0.3 to 0.8 for non-fragmenting disks.  
The lifetime of a disk under such high viscosity is on the order of $10^4$ yr at 
20 au ($t_{\rm visc} = R^2/\alpha c_{\rm s}$). 
Furthermore, non-axisymmetric structures should develop on dynamical time scales 
($\sim$ 9--20 years) at the radii being considered\citemain{lr04}.
Since non-axisymmetric structures are not observed in either gas or dust, 
this implies the disk is stable and has a lower disk-to-stellar 
mass ratio.

\paragraph{Effectiveness of photo-destruction.}

Given a $0.4 \ M_\mathrm{Sun}$ star with an effective temperature 
of 4000 K\citemethods{white04}, a bolometric luminosity $2.7 \ L_\mathrm{Sun}$ 
and a blackbody spectral energy distribution, our modelling predicts that it is 
not possible to substantially photodissociate CO in the disk out to 15 au, 
regardless of the vertical structure or disk's mass distribution. 
For these stellar properties, photodissociation of CO and its isotopologues 
is only efficient near the disk inner edge ($<$ 0.5 au)\citemethods{bruderer13}.  
Furthermore, high-velocity $^{12}$CO 2--1 molecular emission is observed
inside a radius of 15 au but above the disk surface (in the  
outflow)\citemain{bjerkeli2016}.
This suggests that $^{12}$CO molecules are present in the disk inside 15 au.
Fundamental CO ro-vibrational lines have also been detected in both absorption 
and emission toward TMC1A\citemain{herczeg11}, and were shown to originate from 
within 0.2 au.

Following Visser et al.\citemethods{visser09co}, a blackbody of 4000 K with 
the addition of 1 Draine field removes CO up to $A_{\rm v}$ = 2.3. 
Since the photodissociation rate\citemethods{evd06} is 
$k_{\rm diss}\propto\exp\left(-\gamma A_{\rm v}\right)$, where $\gamma$ is 
an attenuation factor and $A_{\rm V}$ is the visual extinction, the amount of 
UV radiation necessary to photodissociate a molecule increases exponentially with 
column density.  
By calculating the vertical $A_{\rm v}$ at different radii, we determine that 
1 Draine field can only dissociate 10$^{-4}$ of the total CO column at 14 au.
This only effectively destroys CO on the disk surface at $>5$ scale heights 
above the disk midplane. To entirely dissociate CO would require $\sim\! 10^4$ Draine 
fields.
X-rays from accretion, meanwhile, tend to produce a warm ionized layer at the surface 
which increases the amount of H$_3^{+}$, which readily destroys CO to form HCO$^{+}$.  
Such chemistry drives the formation of HCO$^{+}$ on the warm surface layers while 
decreasing the CO abundance.  
Nevertheless, even if we adopt a low $^{12}$CO relative abundance of $10^{-6}$, we 
still predict that CO should be observable inside 15 au. 
In either case, for reasonable UV and X-ray intensities, the reduction of 
CO abundance in the disk surface layers is simply not enough to explain 
the observed hole in the molecular emission.

\end{methods}

\begin{addendum}	
	\item[Code Availability] {\it Hyperion} is a public Monte Carlo continuum 
    radiative transfer that can be downloaded from http://www.hyperion-rt.org/. 
    It also contains a routine to calculate the dust opacities. 
    The primary line radiative transfer code is described in Bruderer et al.\citemain{bruderer12}. 
    The code is not public due to the lack of documentation and 
    its non-trivial usage.    
    \item[Data availability] The datasets generated and/or analysed during 
    the current study are available in the ALMA archive, 
    https://almascience.nrao.edu/alma-data/archive, and from 
    the corresponding author.
    
\end{addendum}

\bibliographystylemethods{naturemag}
\bibliographymethods{biblio_methods}


\end{document}